\documentclass{iaus}
\usepackage{graphicx}
\usepackage{graphicx,bm,url}
\usepackage{natbib}

\newcommand{\EQ}{\begin{equation}}
\newcommand{\EN}{\end{equation}}
\newcommand{\EQA}{\begin{eqnarray}}
\newcommand{\ENA}{\end{eqnarray}}
\newcommand{\eq}[1]{(\ref{#1})}

\newcommand{\Fig}[1]{Figure~\ref{#1}}

\newcommand{\bra}[1]{\langle #1\rangle}

{}
{}
{}

{}
{}
\newcommand{\meanEMF}{\overline{\mbox{\boldmath ${\cal E}$}}{}}{}
{}
{}
{}
{}
{}
{}
\newcommand{\meanBB}{\overline{\mbox{\boldmath $B$}}{}}{}
{}
{}
\newcommand{\meanFFf}{\overline{\mbox{\boldmath $F$}}_{\rm f}{}}{}
{}
{}
{}
{}
{}
{}
{}
\newcommand{\meanUU}{\overline{\bm{U}}}

{}
{}
{}

\newcommand{\meanhf}{\overline{h}_{\rm f}}
\newcommand{\meanFFFFf}{\overline{\mbox{\boldmath ${\cal F}$}}_{\rm f}{}}{}

{}

{}
{}

\newcommand{\uu}{\mbox{\boldmath $u$} {}}

\def\bb{\bm{b}}
\newcommand{\BB}{\mbox{\boldmath $B$} {}}

\newcommand{\jj}{\mbox{\boldmath $j$} {}}

\newcommand{\aaaa}{\mbox{\boldmath $a$} {}}

\newcommand{\nab}{\mbox{\boldmath $\nabla$} {}}

\newcommand{\dd}{{\rm d} {}}

\newcommand{\const}{{\rm const}  {}}

\def\ga{\mathrel{\mathchoice {\vcenter{\offinterlineskip\halign{\hfil
$\displaystyle##$\hfil\cr>\cr\sim\cr}}}
{\vcenter{\offinterlineskip\halign{\hfil$\textstyle##$\hfil\cr>\cr\sim\cr}}}
{\vcenter{\offinterlineskip\halign{\hfil$\scriptstyle##$\hfil\cr>\cr\sim\cr}}}
{\vcenter{\offinterlineskip\halign{\hfil$\scriptscriptstyle##$\hfil\cr>\cr\sim\cr}}}}}
\def\Pm{\mbox{\rm Pr}_M}
\def\Rm{\mbox{\rm Re}_M}
\def\Ro{\mbox{\rm Ro}}

\def\Rmc{R_{\rm m,{\rm crit}}}
\def\Rey{\mbox{\rm Re}}

\def\Beq{B_{\rm eq}}

\newcommand{\Mm}{\,{\rm Mm}}

\newcommand{\AU}{\,{\rm AU}}

\newcommand{\yssr}[3]{ #1, {Spa.\ Sci.\ Rev.,} {#2}, #3}
\newcommand{\yapj}[3]{ #1, {ApJ,} {#2}, #3}

\newcommand{\yapjl}[3]{ #1, {ApJL,} {#2}, #3}

\newcommand{\yan}[3]{ #1, {Astron.\ Nachr.,} {#2}, #3}

\newcommand{\yana}[3]{ #1, {A\&A,} {#2}, #3}

\newcommand{\yass}[3]{ #1, {Ap\&SS,} {#2}, #3}
\newcommand{\ygafd}[3]{ #1, {Geophys.\ Astrophys.\ Fluid Dyn.,} {#2}, #3}

\newcommand{\yjfm}[3]{ #1, {J.\ Fluid Mech.,} {#2}, #3}

\newcommand{\ypf}[3]{ #1, {Phys.\ Fluids,} {#2}, #3}

\newcommand{\yprl}[3]{ #1, {Phys.\ Rev.\ Lett.,} {#2}, #3}

\newcommand{\ymn}[3]{ #1, {MNRAS,} {#2}, #3}

\newcommand{\ysci}[3]{ #1, {Science,} {#2}, #3}
\newcommand{\ysph}[3]{ #1, {Solar Phys.,} {#2}, #3}

\newcommand{\ypre}[3]{ #1, {Phys.\ Rev.\ E,} {#2}, #3}

\newcommand{\yspd}[3]{ #1, {Sov.\ Phys.\ Dokl.,} {#2}, #3}

\newcommand{\yjour}[4]{ #1, {#2}, {#3}, #4}

\newcommand{\ybook}[3]{ #1, {#2} (#3)}

\newcommand{\sapj}[2]{ #1, {ApJ}, submitted, arXiv:{#2}}

\title[Non-linear and chaotic dynamo regimes]
{Non-linear and chaotic dynamo regimes}

\author[A.\ Brandenburg] {Axel Brandenburg$^{1,2}$}

\affiliation{
$^1$Nordita, KTH Royal Institute of Technology and Stockholm University,\\
Roslagstullsbacken 23, SE-10691 Stockholm, Sweden\\
$^2$Department of Astronomy, Stockholm University, SE-10691 Stockholm, Sweden\\
}

\pubyear{2013}
\volume{294}
\pagerange{387--398}
\date{$ $Revision: 1.10 $ $}
\jname{Solar and Astrophysical Dynamos and Magnetic Activity}
\editors{A.G. Kosovichev, E.M. de Gouveia Dal Pino \& Y. Yan, eds.}

\begin{document}
\topmargin 2cm

\maketitle

\begin{abstract}
An update is given on the current status of solar and stellar dynamos.
At present, it is still unclear why stellar cycle frequencies increase
with rotation frequency in such a way that their ratio increases with
stellar activity.
The small-scale dynamo is expected to operate in spite of a small
value of the magnetic Prandtl number in stars.
Whether or not the global magnetic activity in stars is a shallow
or deeply rooted phenomenon is another open question.
Progress in demonstrating the presence and importance of
magnetic helicity fluxes in dynamos is briefly reviewed,
and finally the role of nonlocality is emphasized in modeling stellar
dynamos using the mean-field approach.
On the other hand, direct numerical simulations have now come to the
point where the models show solar-like equatorward migration that can be
compared with observations and that need to be understood theoretically.
\keywords{MHD -- turbulence -- Sun: magnetic fields}
\end{abstract}

\firstsection

\section{Introduction}

The objective of IAU Symposium 294 is to provide an update since IAU
Symposium 157 on ``The Cosmic Dynamo'' in Potsdam, 1992.
The title of the present talk reflects the common thinking at that time
that nonlinearity and chaos tend to come together. This is highlighted
by the realization that a simple mean-field dynamo model for poloidal
and toroidal fields needs to be supplemented by a third equation to
produce chaos, and the idea that this third equation is the equation
of magnetic helicity conservation.
These developments have contributed to the unfortunate perception of
mean-field dynamo theory being just as a toy rather than a quantitatively
predictive theory.

With the advent of numerous computer simulations of hydromagnetic
turbulence exhibiting large-scale dynamo action, a new field of
computer astrophysics has emerged where the objective is to understand
simulated dynamos, where one has a chance to resolve all time and length
scales. This approach has helped making mean-field theory quantitatively
reliable and predictable.

Various predictions have emerged and have been tested. Firstly, dynamos
must transport magnetic helicity to escape catastrophic quenching. They
do this through coronal mass ejections and through turbulent exchange
across the equator. The resulting field is bi-helical, with opposite
signs of magnetic helicity at large and small length scales. The signs
depend on the sign of kinetic helicity and on the relative importance of
turbulent diffusion. This has meanwhile been confirmed observationally
using Ulysses data. Secondly, mean-field theory also predicts the
formation of local magnetic flux concentrations as a result of strong
density stratification. Simulations have now confirmed this remarkable
theoretical prediction. This has opened the floor for suggestions that
active regions and sunspots might be shallow phenomena operating near
the surface at some 40 Mm depth.
 
There are also several observations that do not yet have a satisfactory
explanation.
One of them concerns the dependence of the observed cycle period of
late-type stars on their activity and another one is the equatorward
migration of toroidal magnetic flux belts during the solar cycle: is it
caused by meridional circulation, the migratory properties
of the dynamo wave, or something else that we do not know about yet?

\section{Dynamo regimes}

\begin{figure}[t!]\begin{center}
\includegraphics[width=\textwidth]{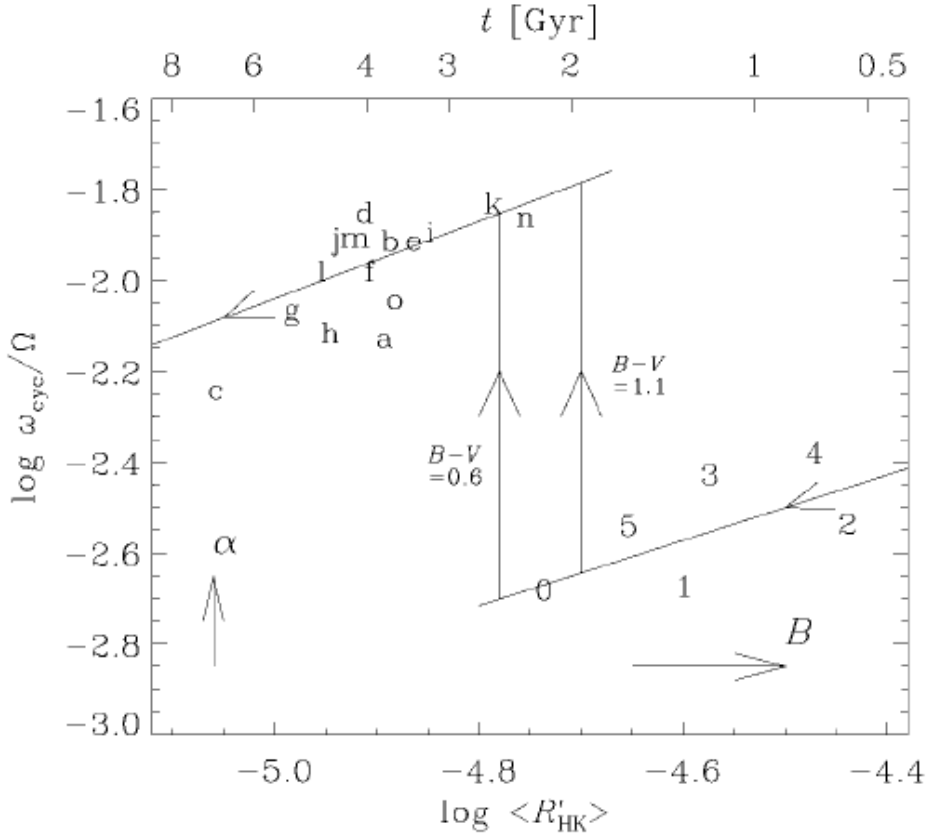}
\end{center}\caption{
BST diagram showing inactive (letters) and active
(numbers) stars on separate branches of normalized
cycle frequency $\omega_{\rm cyc}/\Omega$ as a function of normalized
chromospheric calcium H and K line activity parameter $\bra{R'_{\rm HK}}$.
Approximate age is given on the upper abscissa.
As stars evolve, they move along the lower (active) branch toward the left,
and then (when the color $B-V$ is between 0.6 and 1.1)
jump onto the upper (inactive) branch.
Adapted from \cite{BST98}.
}{}\label{BSTdiagram}\end{figure}

The graph of the ratio of stellar cycle to
rotational frequencies versus magnetic activity or stellar age shows
two branches \citep[][hereafter BST]{BST98}; see \Fig{BSTdiagram}.
These branches correspond to active (A) and inactive (I) stars and are
separated by what is known as the Vaughan--Preston gap.
In addition, there is a third branch of super-active (S) stars \citep{SB99}.
How can we understand the origin of these branches?

The BST diagram is not just a way of representing the non-dimensional
cycle frequency in a two-dimensional diagram, it might represent some
deeper physics.
In a globally quenched $\alpha\Omega$-type dynamo, i.e., a model where only
the smallest wavenumbers have a significant contribution
\citep[as found in simulations; see][]{BRS08,RB12}, the cycle frequency
$\omega_{\rm cyc}$ is proportional to $\sqrt{\alpha\Omega'}$, where
$\alpha$ represents the $\alpha$ effect responsible for reproducing the
large-scale magnetic field \citep{Mof78,KR80} and $\Omega'$ is the
radial differential rotation.
Both are functions of the angular velocity $\Omega$ (which has been subsumed
into a factor $\Omega^q$ with a poorly constrained exponent $q$; see BST).

The essential point is that $\omega_{\rm cyc}$ can be regarded as a proxy
of $\alpha$.
Furthermore, the angular velocity normalized by the turnover time $\tau$,
which gives the inverse Rossby number $\Ro^{-1}=2\Omega\tau$,
can be regarded as a measure of the non-dimensional magnetic field
strength, $B/\Beq$, where $\Beq$ is the equipartition field strength.
This is a relationship that is {\it independent} of the existence of
different branches, i.e., inactive and active stars lie on the same curve
(BST).
We can therefore imagine the BST diagram being really a
representation of $\alpha(B)$,
and that the two rising branches describe therefore an anti-quenching of
$\alpha$ with $B/\Beq$.
Indications of this, and a corresponding anti-quenching of the
turbulent magnetic diffusivity, have been found in simulations of
magneto-buoyancy \citep{Chatter11}.
On the other hand, for faster rotation there is a third branch of
super-active stars \citep{SB99}, for which our proxy of $\alpha$
declines with that of $B$.

These considerations are still as exciting today as they were back
in 1998, but now we have realistic global simulations of convection
that reflect a qualitative leap from earlier work in that we now find
for the first time cyclic large-scale dynamo action with equatorward
migrating activity belts \citep{KMB12}.
One needs to check, however, whether perhaps all of the dynamo solutions
obtained so far are representative of the superactive branch, and that
the physics behind the active and inactive ones remains still to be
discovered.
To make progress in understanding the different modes of cyclic stellar
activity, one also needs to analyze why those models produce long cycle
periods and equatorward migration, both of which are also seen in the Sun,
but are theoretically not understood;
is it related to the possible dominance of the magnetic $\alpha$ effect
\citep{PFL} (which is inversely proportional to the density and therefore
important near the surface), to the tensorial structure of $\alpha$
($\alpha_{ij}$) and turbulent diffusion ($\eta_{ijk}$),
to meridional circulation, or to subtleties in the differential rotation?
Such understanding should be accomplished by deriving and solving suitable
mean-field models that reproduce the behavior seen in DNS and
Large Eddy Simulations (LES) of the Sun.

It is unlikely that differences in the cycle period are the only criterion
distinguishing stars on the two branches of the evolutionary BST diagram.
Surface magnetic field structures as well as their spatio-temporal correlations
are now becoming accessible to detailed scrutiny.
Quantifying the nature of magnetic fields
using observed correlations among the Stokes parameters might help
to distinguish different types of behaviors and to associate them with
different branches in the BST diagram, which may reflect different
underlying dynamo modes.
Progress can be made by considering turbulent dynamo simulations at different
rotation rates, as has recently been done by \cite{KMCWB13}.
We return to this issue in our conclusions and turn attention to recent
simulation results that concern the magnetic surface activity of the Sun,
such as small-scale or local dynamos and the evidence of helical magnetic
fields from the global dynamo.

\section{Small-scale and local dynamos}

In recent years the action of two separate dynamos in the Sun has become
popular; one that governs the 11 year cycle and one that produces the
small-scale field of the quiet Sun \citep{Cat99,CEW03,VS07}.
This idea is supported by the fact that the observed small-scale field
of the Sun is essentially uncorrelated with that of the 11 year cycle
\citep{Lit02,IT09,DSS10,Aur10,Ste12}.

A potential problem is the fact that the critical magnetic Reynolds
number $\Rmc$ grows larger as one decreases the value of the magnetic
Prandtl number, $\Pm=\nu/\eta$, to more realistic values \citep{Scheko05}.
Early work of \cite{RK97} did already predict an increased value of
$\Rmc$ in the limit of small values of $\Pm$.
\cite{BC04} argue that the reason for an increased value of $\Rmc$ is
connected with the ``roughness'' of the velocity field.
\cite{Iska07} found that $\Rmc$ has a local maximum at $\Pm=0.1$,
and that it decreases again as $\Pm$ is decreased further.
The reason for this is that near $\Pm=0.1$ the resistive wavenumber is
about 10 times smaller than the viscous one and thus right within the
``bottleneck'' where the spectrum is even shallower than in the rest
of the inertial range, with a local scaling exponent that
corresponds to turbulence that is in this regime rougher still,
explaining thus the apparent divergence of $\Rmc$.

\begin{figure}[t!]\begin{center}
\includegraphics[width=.8\columnwidth]{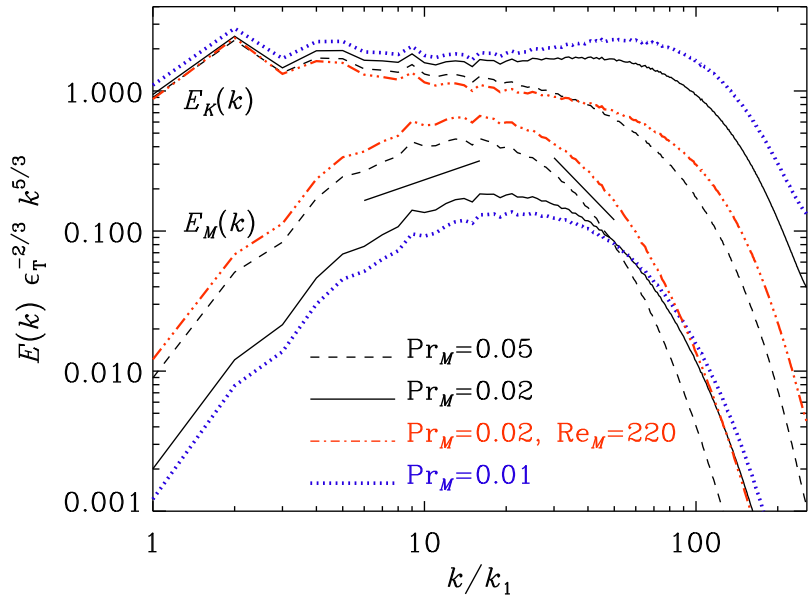}
\end{center}\caption[]{
Compensated kinetic and magnetic energy spectra for runs
with $\Pm=0.05$, $\Pm=0.02$, and $\Pm=0.01$ for $\Rm\approx160$
as well as one run with $\Pm=0.02$ and $\Rm\approx220$.
The resolution is in all cases $512^3$ mesh points.
The two short straight lines give, for comparison, the slopes
$2/3$ (corresponding to a $k^{-1}$ spectrum for $k<20k_1$) and
$-2$ (corresponding to a $k^{-11/3}$ spectrum for $k>20k_1$).
Adapted from \cite{B11}.
}\label{pspec2}\end{figure}

In the nonlinear regime the magnetic field affects the flow in
such a way that the bottleneck effect tends to be suppressed,
so the divergence in the roughness disappears and there is a smooth
dependence of the saturation field
strength on the value of $\Pm$; see \cite{B11} for details.
In \Fig{pspec2}, we show spectra compensated with $\epsilon^{-2/3}k^{5/3}$.
For $\Pm=0.02$ and 0.01, the kinetic energy spectra show a clear
bottleneck effect, i.e., there is a weak uprise of the compensated spectra
toward the dissipative subrange \citep{Fal94,Kan03,Dob03}.
The compensated magnetic energy spectra peak around $k=20k_1$,
where $k_1=2\pi/L$ is the smallest wavenumber in a domain of size $L$.
Both toward larger and smaller values of $k$ there is no clear
power-law behavior. The slopes of the $k^{-11/3}$ spectrum
of \cite{Gol60} and \cite{Mof61} and the scale-invariant $k^{-1}$ spectrum
\citep{RS82,KR94,KMR96} are shown for comparison.

We recall that we have used here the strategy of generating
low-$\Pm$ solutions by gradually decreasing $\nu$, and hence
increasing the value of $\Rey$.
As in the case of helical dynamos \citep{B09}, the fact that
a turbulent self-consistently generated magnetic field is present
helps reaching these low-$\Pm$ solutions.
However, the presence of the magnetic field also modifies the
kinetic energy spectrum and makes it decline slightly more steeply
than in the absence of a magnetic field; see \Fig{pspec2}.
This suggests that the velocity field would be less rough
than in the corresponding case without magnetic fields.
Following the reasoning of \cite{BC04}, this should make
the dynamo more easily excited than in the kinematic case
with an infinitesimally weak magnetic field.
In other words, there is the possibility of a subcritical bifurcation
where the dynamo requires a significantly larger value of $\Pm$ to
bifurcate from the trivial $\BB=\bm{0}$ solution than the value
needed to sustain a saturated dynamo.

\begin{figure}[t!]\begin{center}
\includegraphics[height=4cm]{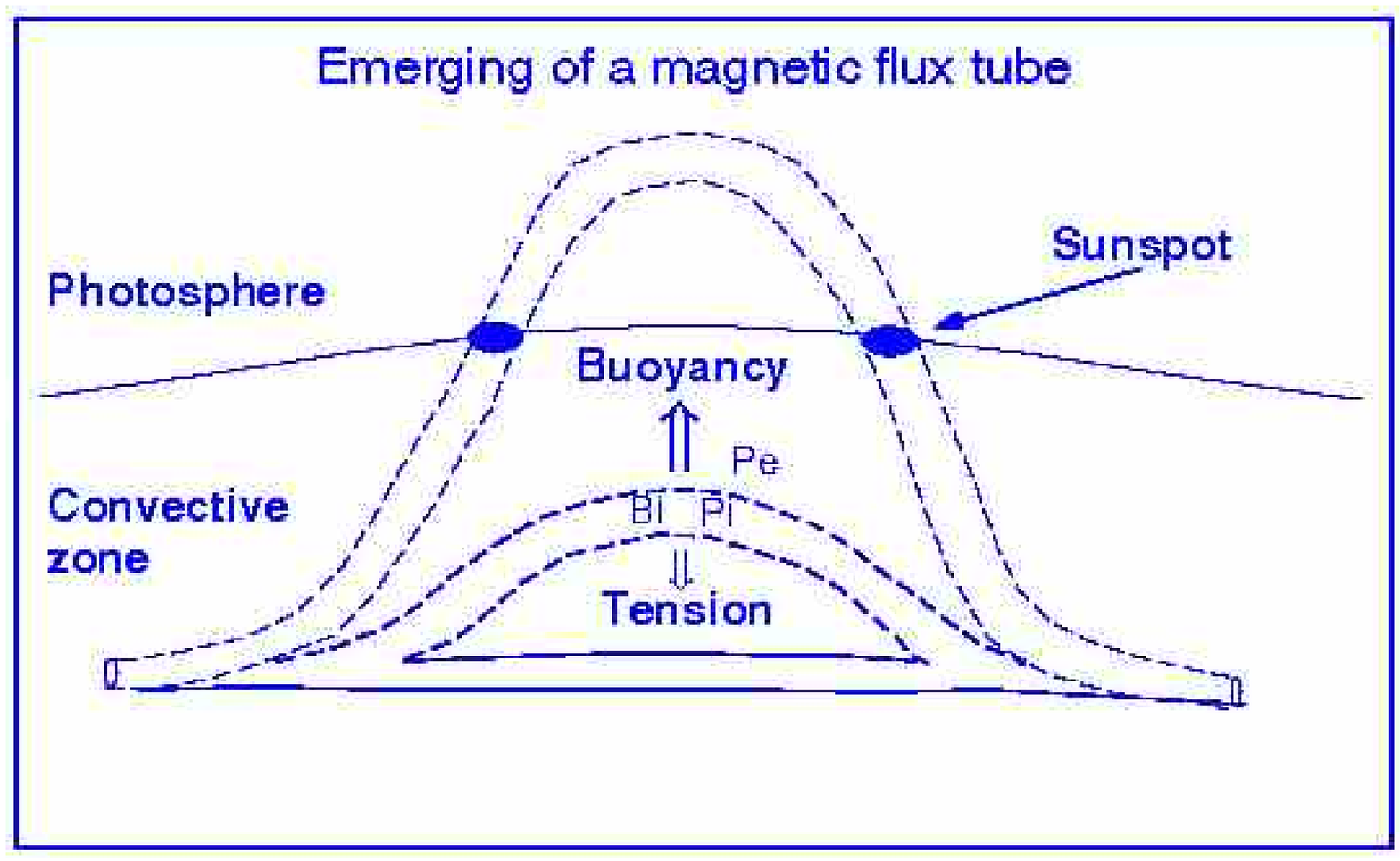}
\includegraphics[height=4cm]{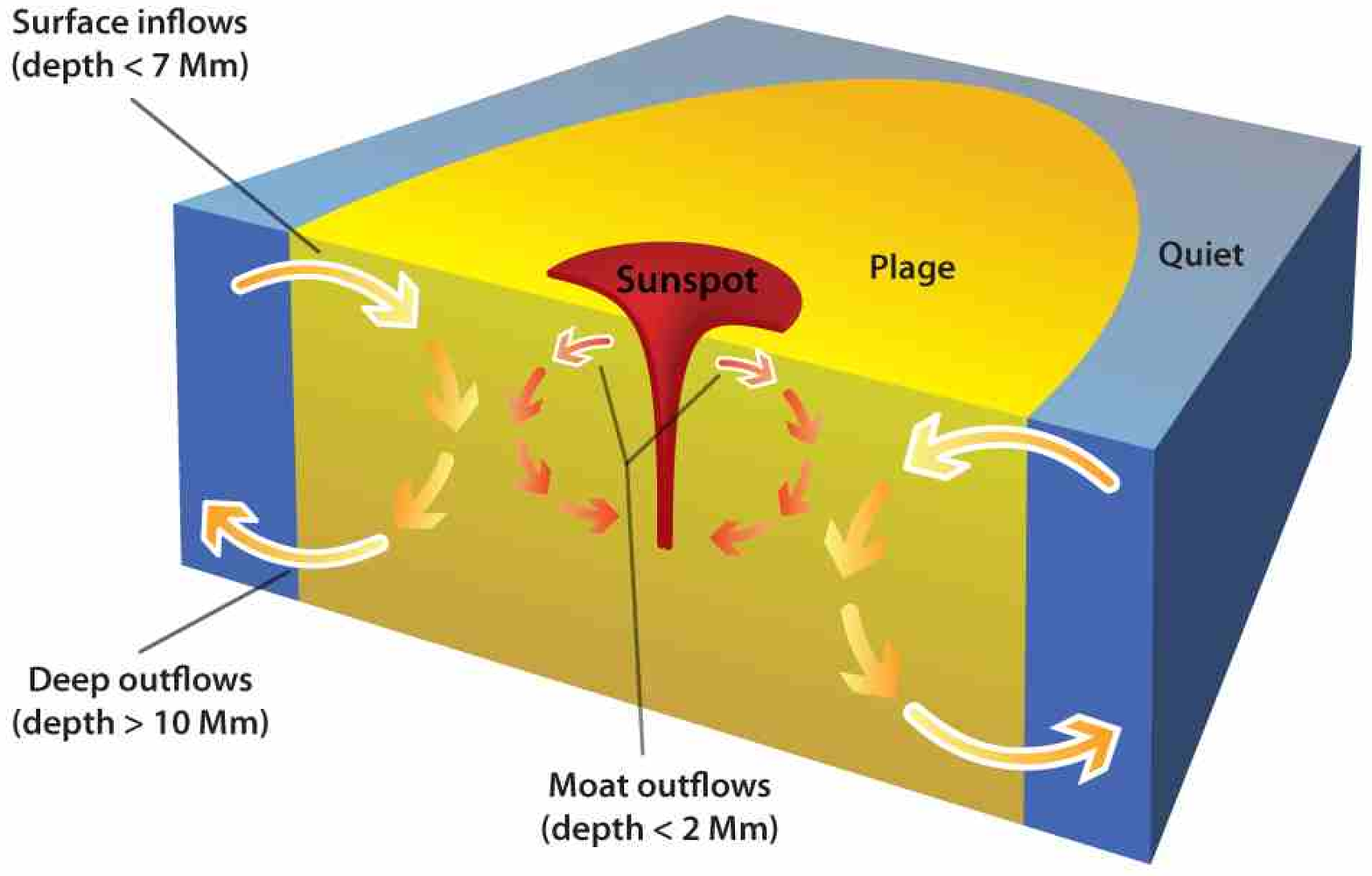}
\end{center}\caption[]{\small
{\it Left}: rising flux tube piercing the surface to form a pair of sunspots
(taken from {\texttt http://www.lund.irf.se/helioshome/fluxtube.gif}).
{\it Right}: sunspot with surrounding flow field suggested from local
helioseismology.
Adapted from \cite{Hindman}.
\normalsize}\label{fluxtube}\end{figure}

\section{Solar surface activity}

The theory of stellar structure explains that the outer $200\Mm$ of the
Sun's radius are convectively unstable, resulting in fully developed
turbulent convection.
Numerical simulations of turbulent flows predict that part of the convective
kinetic energy is converted to magnetic energy through dynamo action.
If we did not have observations, would we have predicted that the
Sun's magnetic field would choose to manifest itself in the form of spots?
The answer might well be yes, but perhaps not for the reasons offered in
text books.
Standard thinking focuses on the tachocline, which is a strong shear layer
at the bottom of the convection zone.
Strong shear can produce a strong magnetic field in
the form of thin flux tubes \citep{Cline,GK11}.
The magnetic pressure in these tubes expels gas, and so, being less
dense than their surrounding, they rise.
If a segment of a tube pierces the surface of the Sun, the footpoints of
the resulting arch appear as sunspot pairs of opposite polarity (as the
magnetic field in the tube has a definite direction; see \Fig{fluxtube}).
Simulations, on the other hand, predict turbulent magnetic fields
with a diffuse large-scale component throughout the convection zone
\citep{BBBMT10,KKBMT10,GCS10}, and this scenario can also reproduce the
observed bipolar spots at the surface \citep{B05}.

\begin{figure}[t!]\begin{center}
\includegraphics[height=4cm]{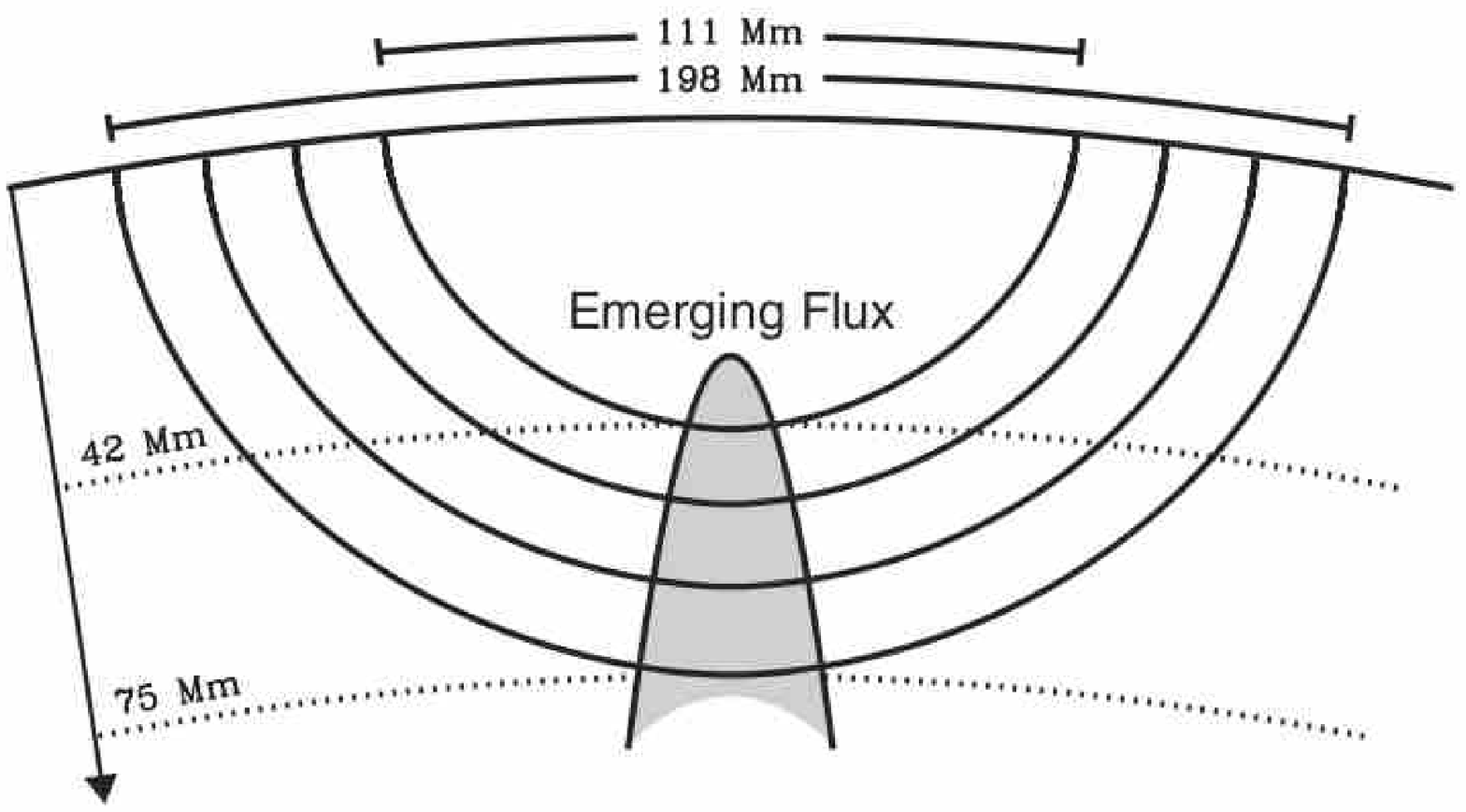}
\includegraphics[height=4.5cm]{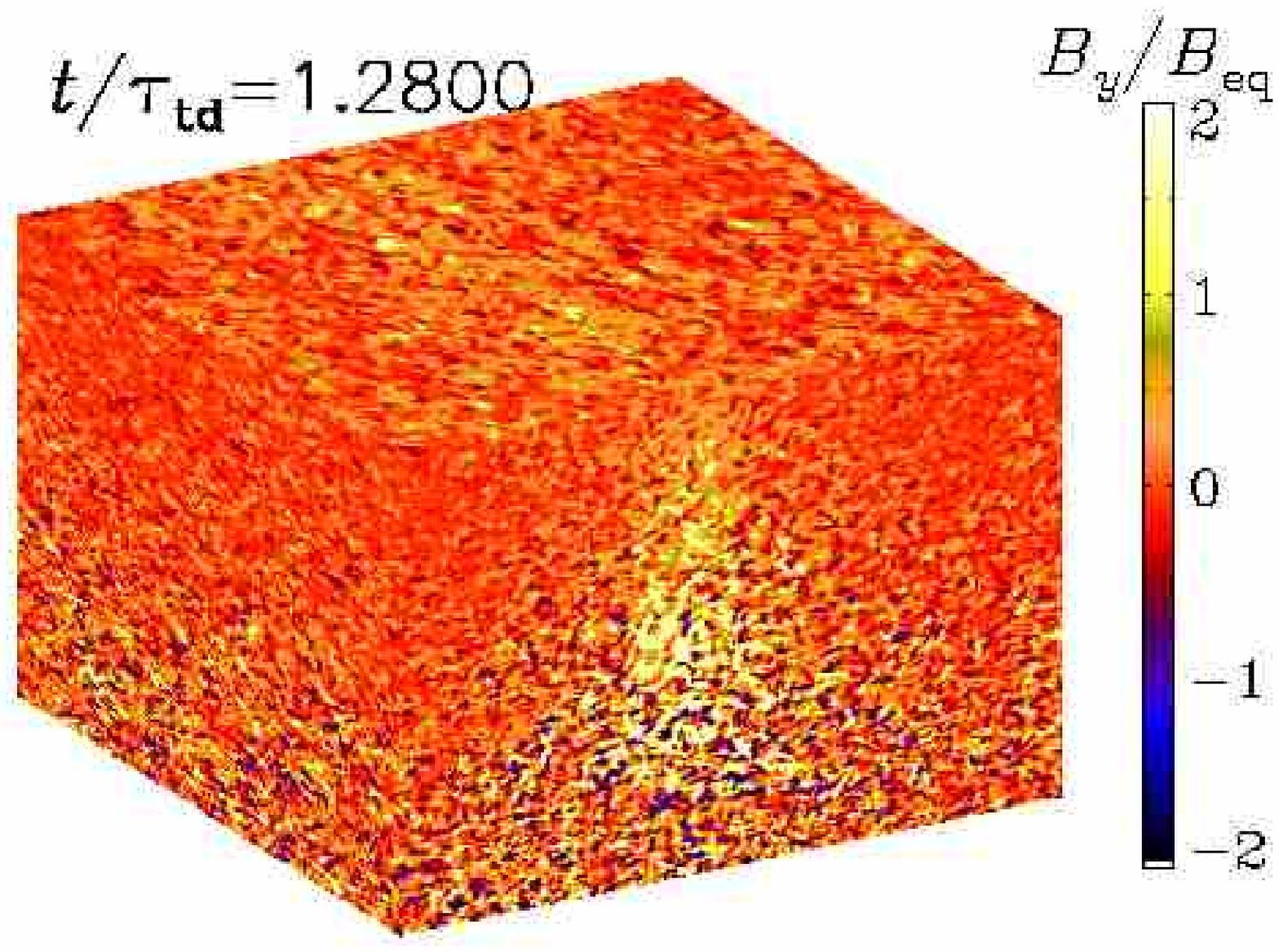}
\end{center}\caption[]{\small
{\it Left}: sketch illustrating the detection of subsurface magnetic
activity via local helioseismology.
Acoustic ray paths are bent back up again because of higher sound speed
at greater depth (lower turning points between 42 and 75 Mm are shown).
Travel-time anomalies allow detection of emergent flux (sketched in gray)
near those turning points.
{\it Right}: visualization of the magnetic field on the periphery of the
computational domain as obtained from NEMPI.
Light-yellow regions indicate enhanced flux in a region reminiscent of
that implied by local helioseismology ({\it left}).
Adapted from \cite{Ilonidis} [left] and \cite{KBKMR12} [right].
\normalsize}\label{Ilonidis}\end{figure}

It might become possible to use local helioseismology
to distinguish between the scenarios sketched in the left
and right hand panels of \Fig{fluxtube}.
Unlike global helioseismology,
local helioseismology is an advanced technique that
uses correlations of measured Doppler
shifts at the solar surface for different time intervals
corresponding to sound travel times for rays down to a
given depth, as is seen in the left-hand panel of \Fig{Ilonidis}.
This technique can provide detailed information on the structure of
magnetic fields \citep{Ilonidis} nearby and even inside a sunspot
\citep{Kos09}.
In a particular case, some type of local activity has been detected at a depth
of $\sim60\Mm$, which corresponds to 1/3 of the depth of the convective zone.
If this was caused by a rising flux tube, as sketched in \Fig{Ilonidis},
one would have expected a wider elongated feature.
On the other hand, the observed activity might correspond
to signatures of magnetic structures formed by the so-called
negative effective magnetic pressure instability \citep[NEMPI,][]{BKKMR11}.

\begin{figure}[t!]\begin{center}
\includegraphics[width=.8\columnwidth]{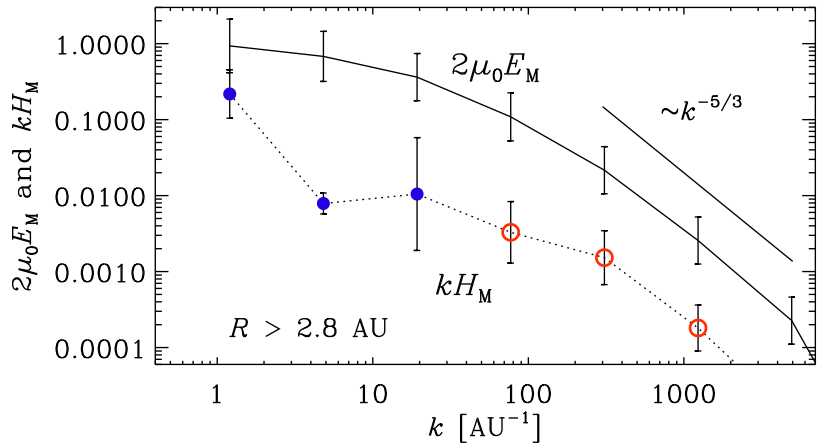}
\end{center}\caption[]{
Magnetic energy and helicity spectra, $2\mu_0 E_{\rm M}(k)$
and $kH_{\rm M}(k)$, respectively, for two separate distance intervals.
Furthermore, both spectra are scaled by $4\pi R^2$ before averaging
within each distance interval above $2.8\AU$.
Filled and open symbols denote negative and positive values of
$H_{\rm M}(k)$, respectively.
Adapted from \cite{BSBG11}.
}\label{rmeanhel4b}\end{figure}

\begin{figure}[t!]\begin{center}
\includegraphics[height=3.9cm]{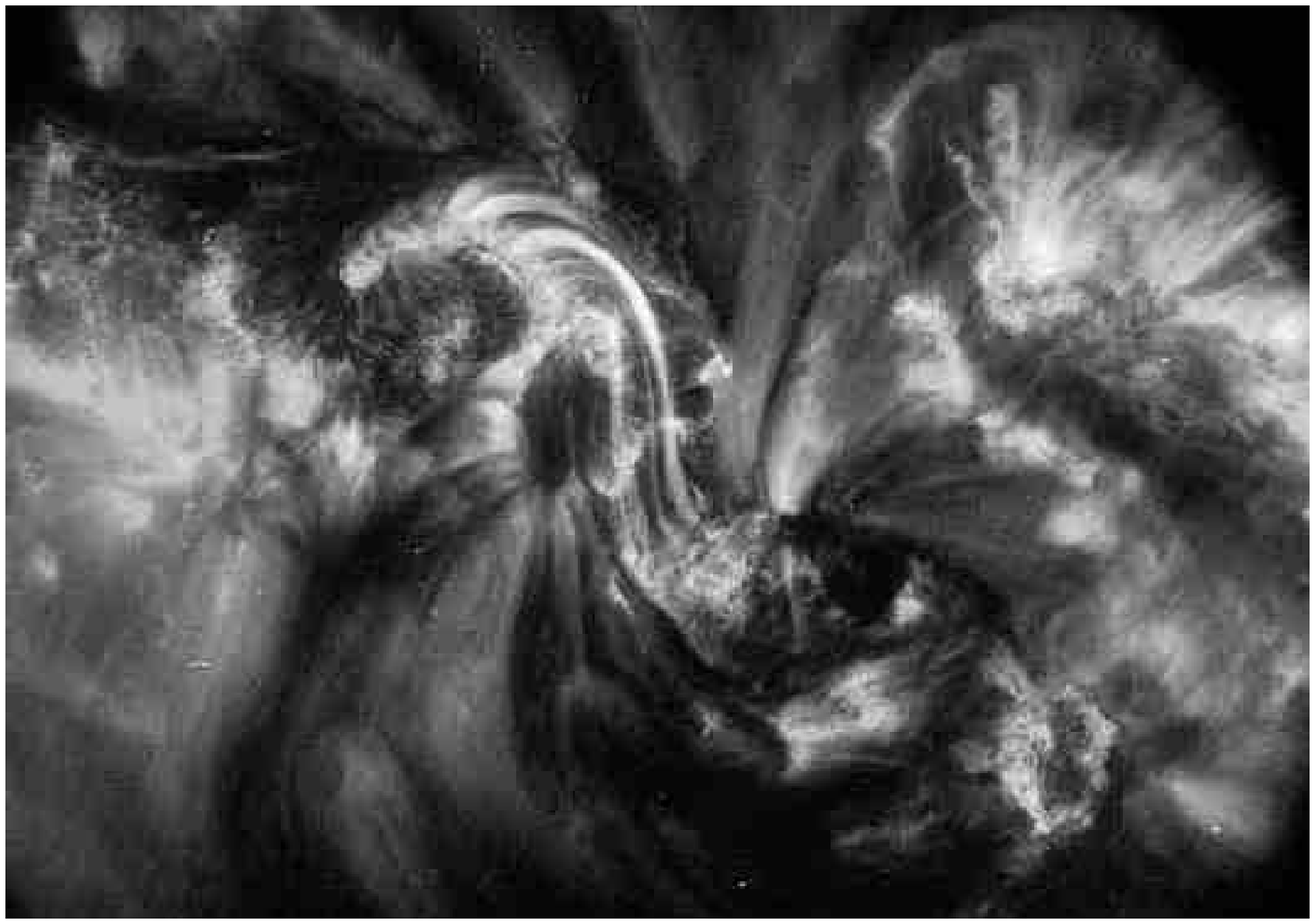}
\includegraphics[height=3.9cm]{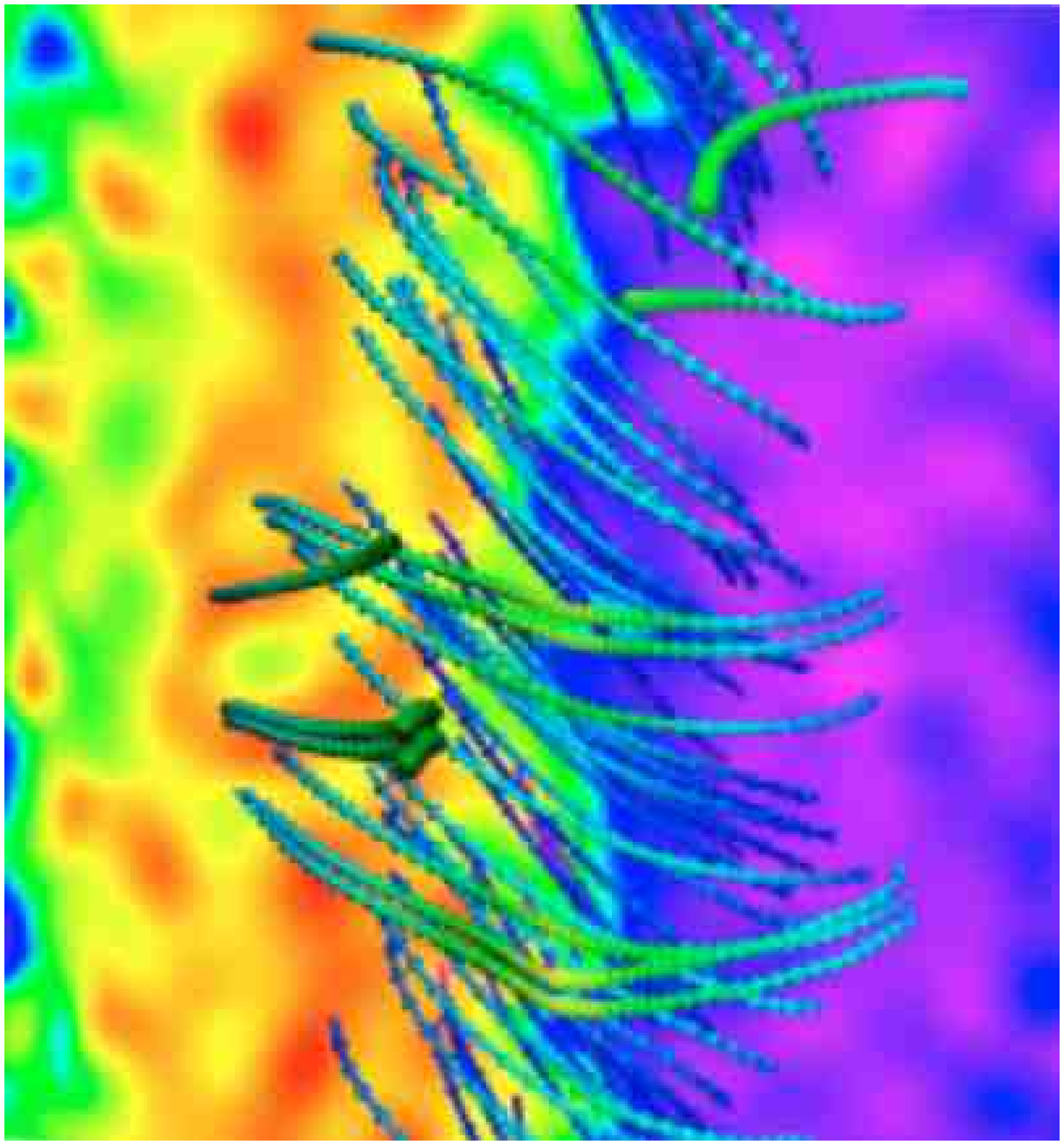}
\includegraphics[height=3.9cm]{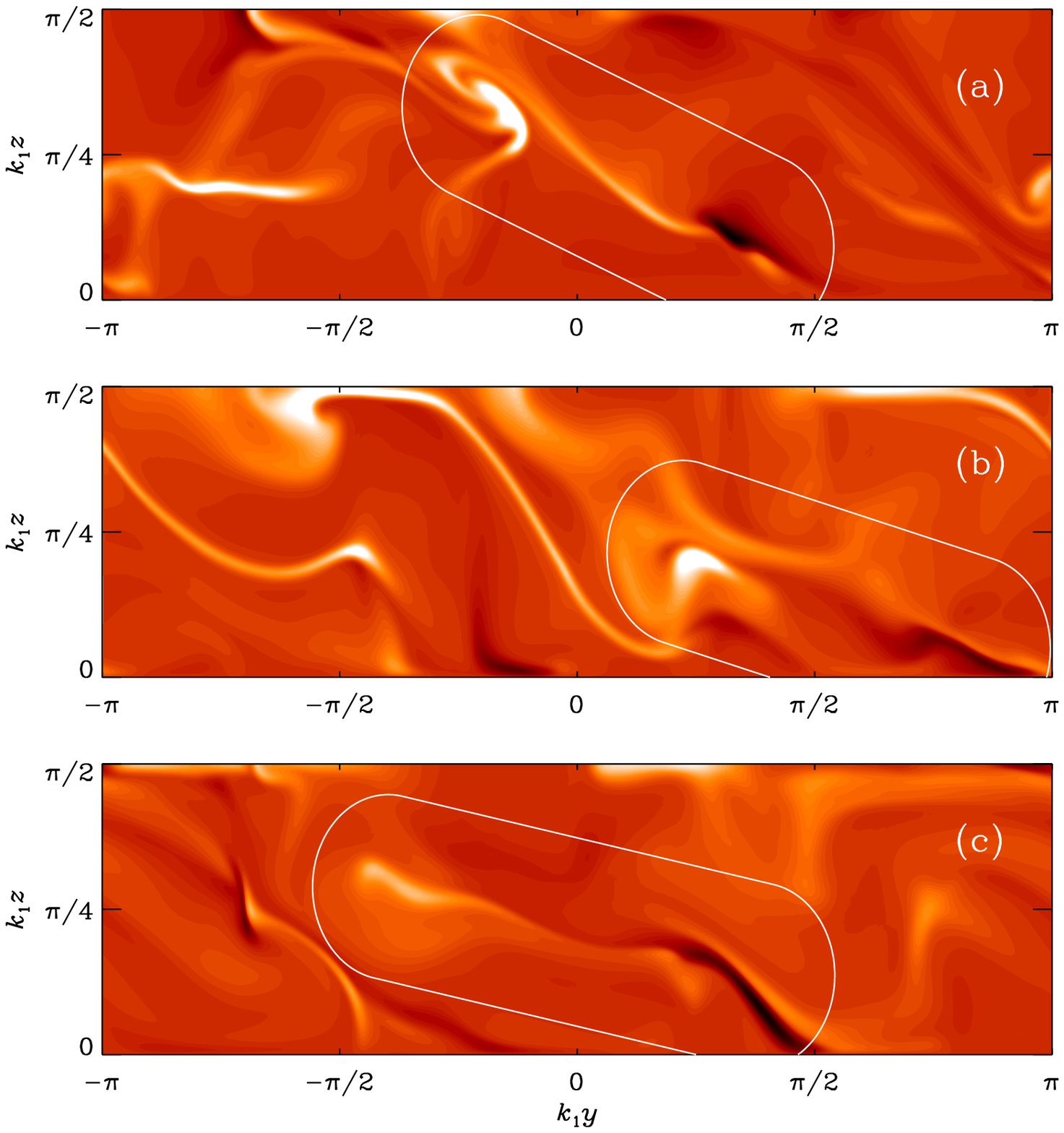}
\end{center}\caption[]{\small
{\it Left}: twisted magnetic field lines from a bipolar region on the
Sun, as seen in X-rays \cite[adapted from][]{gibson}.
{\it Middle}: twisted magnetic field lines from a self-consistently
generated bipolar sheet \citep{WB10}, not a bipolar region.
Here the field is generated by a dynamo without shear.
{\it Right}: bipolar regions as seen in simulations with shear \citep{B05}.
Light (dark) shades correspond to positive (negative) line of sight
magnetic field.
Adapted from \cite{gibson} [left], \cite{WB10} [middle], and
\cite{B05} [right].
\normalsize}\label{gibson}\end{figure}

Coronal mass ejections play a major role
in shedding small-scale magnetic helicity from the dynamo to alleviate
an otherwise catastrophic quenching of the dynamo \citep{BB03}.
Meanwhile, models have made contact with unexpected phenomena
taking place in the solar wind.
A striking example is the sign reversal of small-scale magnetic helicity
away from the Sun.
This surprising result was first obtained by analyzing data from the
{\it Ulysses} spacecraft \citep{BSBG11}, see \Fig{rmeanhel4b},
but the interpretation was
greatly aided by similar results from simulations of \cite{WBM11}.
It now seems that the reason for this is an essentially turbulent-diffusive
transport down the local gradient of magnetic magnetic helicity density --
even in the wind \citep{WBM12}.
While this work has focussed on parameter studies
exploring the conditions for plasmoid ejections from helically
forced turbulence as well as rotating convection, the physical
realism of the model remained poor.
The density contrast between dynamo region and corona is much bigger in
reality, see for example \cite{Pinto}.
Significant improvements are possible with only modest increase of
numerical resolution, as has been shown by \cite{BP11} using
{\sc Pencil Code} simulations with a realistic setup.
One may envisage important follow-up diagnostics by producing
visualizations of helical magnetic fields in the corona
(see the left-hand panel of \Fig{gibson})
and to compute cases in which the field is generated either self-consistently
by a dynamo beneath the surface, as in \cite{WBM11,WKMB12}, or the field
is injected as a twisted flux tube in a deeper layer and let to
emerge at the surface.
Simulations without shear have successfully produced twisted magnetic
field lines from a self-consistently generated bipolar sheet (see middle
panel of \Fig{gibson}), but this has not yet been attempted in simulations
where more localized bipolar regions are produced.
An example of the formation of such regions has been seen in dynamo
simulations with strong shear \citep{B05} leading to the occasional
formation of bipolar regions when opposite polarities can be drawn apart
by latitudinal differential rotation; see the right-hand panel of \Fig{gibson}.
Observational evidence for such a process has been provided by \cite{KS08}.

\begin{figure}[t!]\begin{center}
\includegraphics[width=.8\textwidth]{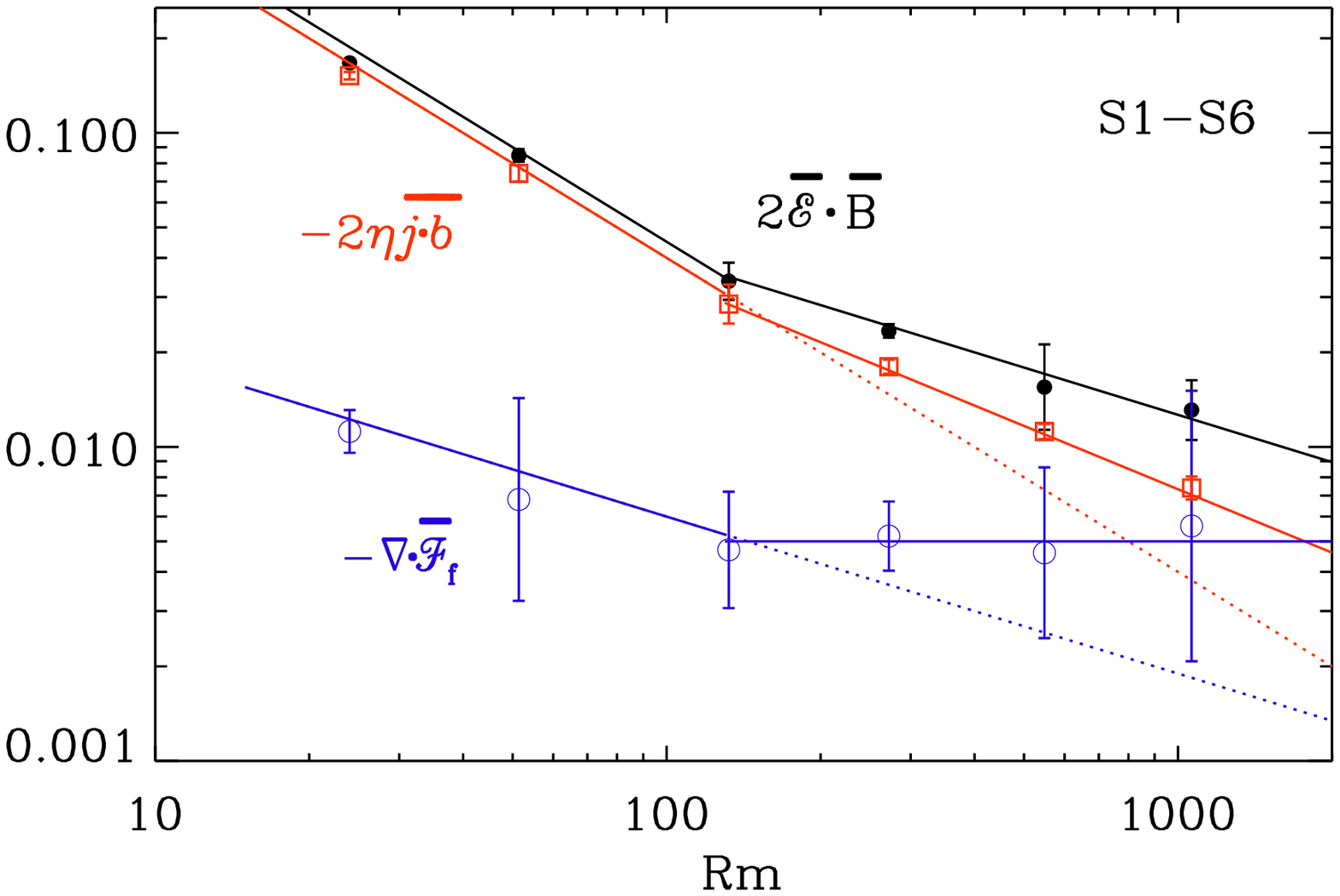}
\end{center}\caption[]{
Scaling properties of the vertical slopes of $2\meanEMF\cdot\meanBB$,
$-2\eta\mu_0\,\overline{\jj\cdot\bb}$, and $-\nab\cdot\meanFFf$.
The three quantities vary approximately linearly with $z$, so
the three labels indicate their non-dimensional values at $k_1z=1$.
Adapted from \cite{DSGB13}.
}\label{presults2b}\end{figure}

Recent work using a simple model with a galactic wind has shown, for the
first time, that shedding magnetic helicity by fluxes may indeed be possible.
We recall that the evolution equation for the mean magnetic helicity density
of fluctuating magnetic fields, $\meanhf=\overline{\aaaa\cdot\bb}$, is
\begin{equation}
{\partial\meanhf\over\partial t}=
-2\meanEMF\cdot\meanBB-2\eta\mu_0\,\overline{\jj\cdot\bb}
-\nab\cdot\meanFFFFf,
\end{equation}
where we allow two contributions to the flux of magnetic helicity from the
fluctuating field $\meanFFFFf$: an advective flux due to the wind,
$\meanFFFFf^{\rm w}=\meanhf\meanUU_{\rm w}$,
and a turbulent--diffusive flux due to turbulence,
modelled here by a Fickian diffusion term down the gradient
of $\meanhf$, i.e., $\meanFFFFf^{\rm diff}=-\kappa_h\nab\meanhf$.
Here, $\meanEMF=\overline{\uu\times\bb}$ is the electromotive force
of the fluctuating field.
The scaling of the terms on the right-hand side with $\Rm$ has been
considered before by \cite{MCCTB10} and \cite{HB10}.

In \Fig{presults2b} we show the basic result of \cite{DSGB13}.
As it turns out, below $\Rm=100$ the $2\eta\mu_0\overline{\jj\cdot\bb}$ term
dominates over $\nab\cdot\meanFFFFf$, but because of the different scalings
(slopes being $-1$ and $-1/2$, respectively), the $\nab\cdot\meanFFFFf$ term is
expected to become dominant for larger values of $\Rm$ (about 3000).
Surprisingly, however, $\nab\cdot\meanFFFFf$ becomes approximately constant
for $\Rm\ga100$ and $2\eta\mu_0\overline{\jj\cdot\bb}$ shows now a shallower
scaling (slope $-1/2$).
This means that that the two curves would still cross at a similar value.
Our data suggest, however, that $\nab\cdot\meanFFFFf$ may even rise slightly,
so the crossing point is now closer to $\Rm=1000$.

\begin{figure}[t!]\begin{center}
\includegraphics[width=.8\textwidth]{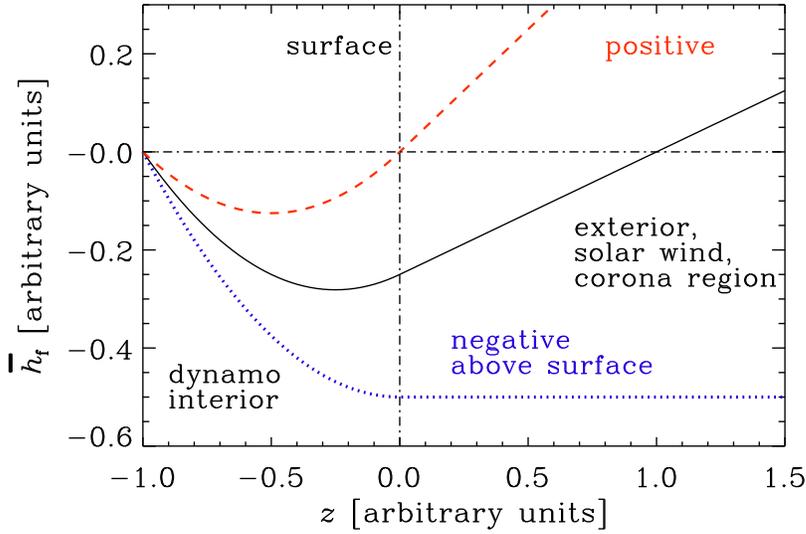}
\end{center}\caption[]{
Sketch showing possible solutions $\meanhf(z)$ (upper panel)
with $S=\const=-1$ in $z<0$ and $S=0$ in $z>0$.
The red (dashed) and black (solid) lines show solutions for which the
magnetic helicity flux ($-\kappa_h\dd\meanhf/\dd z$, see lower panel)
is negative in the exterior.
This corresponds to the case observed in the Sun.
The blue (dotted) line shows the case, where the magnetic helicity
flux is zero above the surface and therefore do not reverse the sign of
$\meanhf(z)$ in the exterior.
Adapted from \cite{WBM12}.
}\label{psketch}\end{figure}

We have mentioned above some surprising behavior that has been noticed in
connection with the small-scale magnetic helicity flux in the solar wind.
Naively, if negative magnetic helicity from small-scale fields
is ejected from the northern hemisphere, one would expect to find
negative magnetic helicity at small scales anywhere in the exterior.
However, if a significant part of this wind is caused by a diffusive
magnetic helicity flux, this assumption might be wrong and the sign
changes such that the small-scale magnetic helicity becomes positive
some distance away from the dynamo regime.
In \Fig{psketch} we reproduce in graphical form the explanation offered
by \cite{WBM12}.

\begin{figure}[t!]\begin{center}
\includegraphics[width=0.58\textwidth]{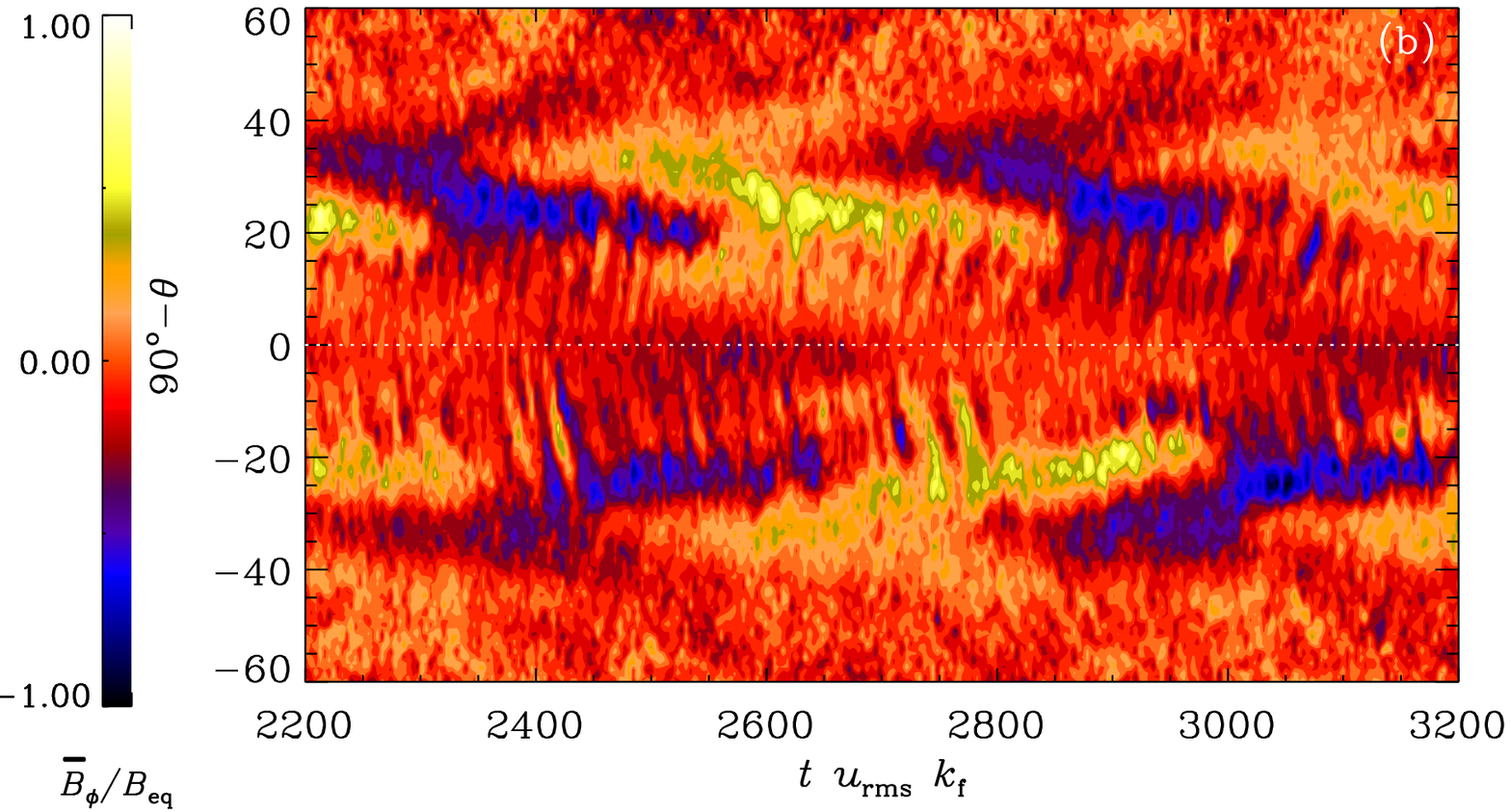}
\includegraphics[width=0.41\textwidth]{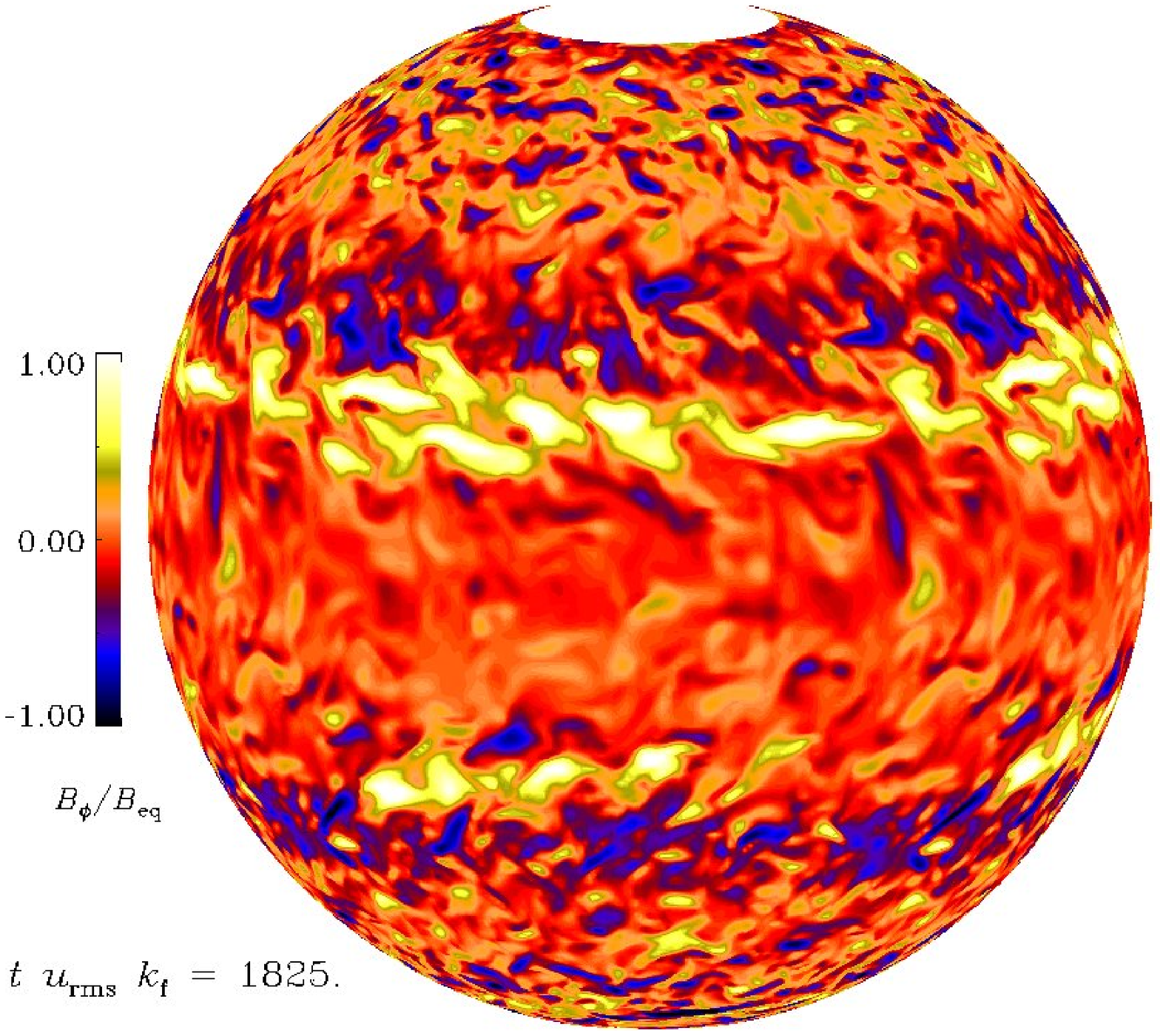}
\end{center}\caption{{\em Left}: azimuthally averaged toroidal magnetic field
as a function of time (in turnover times) and latitude
(clipped between $\pm60^\circ$).
Note that on both sides of the equator ($90^\circ-\theta=\pm25^\circ$),
positive (yellow) and negative (blue) magnetic fields move equatorward,
but the northern and southern hemispheres are slightly phase shifted
relative to each other.
{\em Right}: Snapshot of the toroidal magnetic field $B_\phi$ at $r=0.98$
outer radii.
Adapted from \cite{KMB12}.
}\label{equatorward}\end{figure}

\section{Conclusions and further remarks}

In this review we have put emphasis on the appearance of magnetic helicity
at and above the surface of the dynamo.
Other important diagnostics may come from local helioseismology to distinguish
between shallow and deeply rooted dynamo scenarios.
As mentioned above, simulations by various groups all produce distributed
dynamo action where the magnetic field is present throughout the
convection zone.

A major breakthrough has been achieved through the recent finding of
equatorward migration of magnetic activity belts
in the course of the cycle \citep{KMB12}; see \Fig{equatorward}.
These results are robust and have now been reproduced in extended simulations
that include a simplified model of an outer corona \citep{WKMB12}.
Interestingly, the convection simulations of other groups produce cycles
only at rotation speeds that exceed those of the present Sun by a factor of
3--5 \citep{BMBBT11}; see also \cite{Racine} for recent cyclic models
at solar rotation speeds.
Both lower and higher rotation speeds give, for example, different
directions of the dynamo wave \citep{KMB12}.
Different rotation speeds correspond to different stellar ages
(from 0.5 to 8 gigayears for rotation periods from 10 to 40 days),
because magnetically active
stars all have a wind and are subject to magnetic braking \citep{Sku72}.
In addition, all simulations are subject to systematic ``errors'' in
that they poorly represent the small scales and emulate in that way an
effective turbulent viscosity and magnetic diffusivity that is larger
than in reality; see the corresponding discussion in Sect.~4.3.2 of
\cite{BKKR12} in another context.
In future simulations, it will therefore be essential to explore the range
of possibilities by including stellar age as an additional dimension of the
parameter space.

In future work it will be important to understand the results of simulations
using simpler mean-field models.
A potential problem is the fact that the turbulent eddies often have sizes
comparable with the size of the domain.
In that case, scale separation in space or time is poor and the mean-field
$\alpha$ effect and turbulent diffusivity have to be replaced by integral
kernels by which the dependence of the mean electromotive force on the
mean magnetic field becomes nonlocal.

\begin{figure}[t!]
\centering
\includegraphics[height=6.6cm]{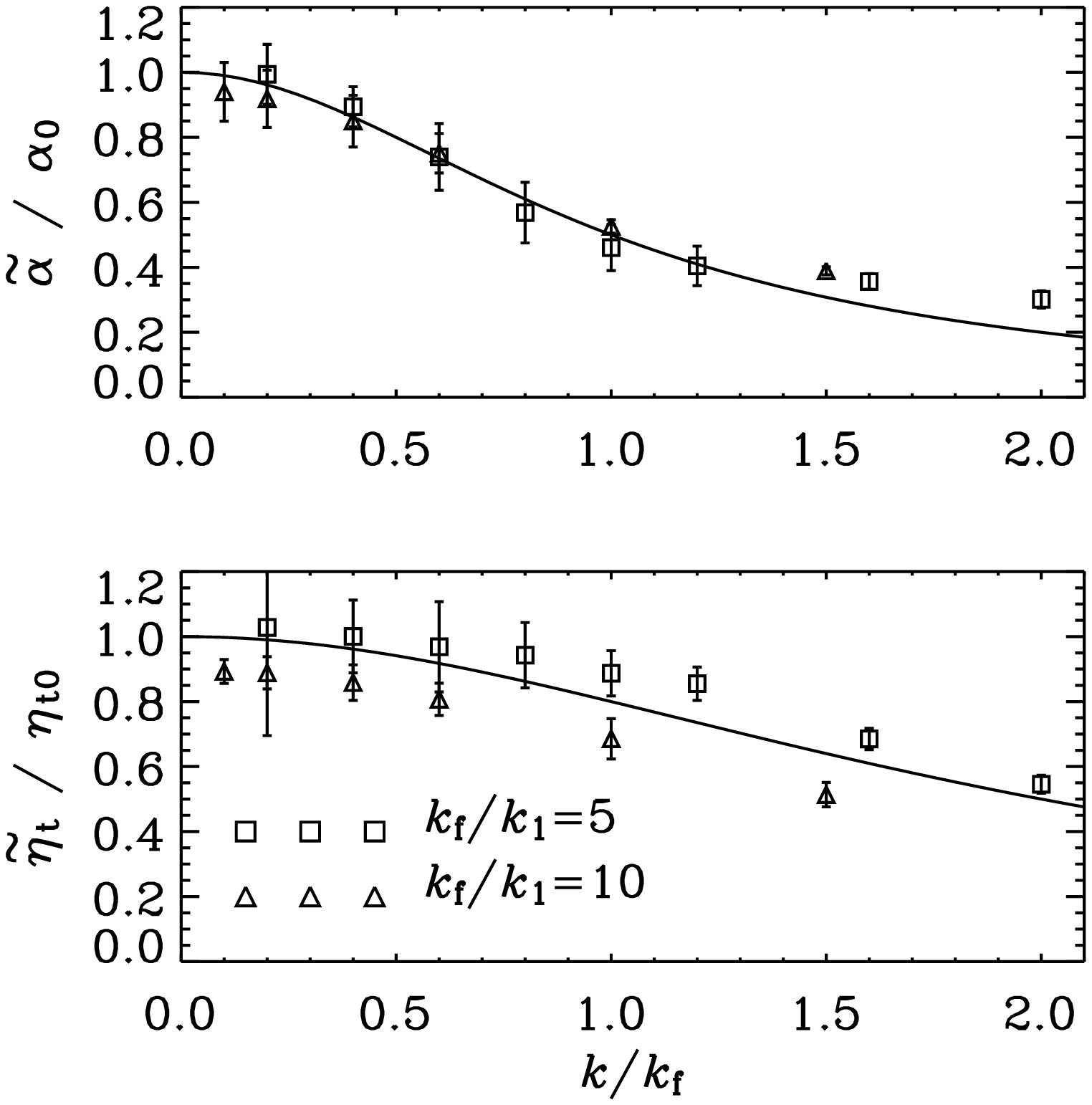}
\includegraphics[height=6.6cm]{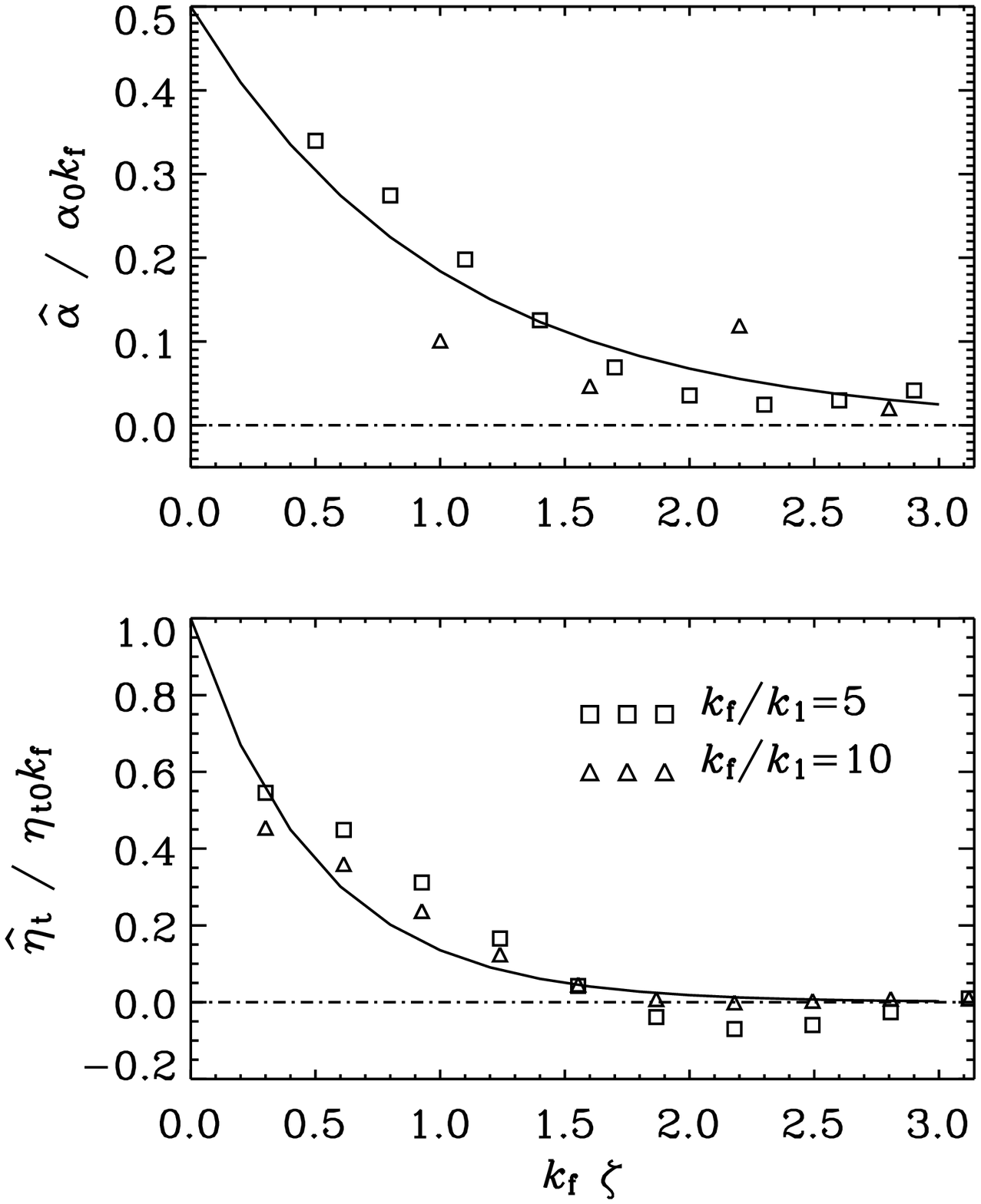}
\caption{
{\it Left}:
Dependences of the normalized $\tilde\alpha$ and $\tilde\eta_{\rm t}$
on the normalized wavenumber $k/k_{\rm f}$
for isotropic turbulence forced at wavenumbers $k_{\rm f}/k_1=5$ with $\Rm=10$ (squares)
and $k_{\rm f}/k_1=10$ with $\Rm=3.5$ (triangles).
The solid lines give the Lorentzian fits \eq{KernelsTurb}.
{\it Right}:
Normalized integral kernels $\hat\alpha$ and $\hat\eta_{\rm t}$ versus
$k_{\rm f}\zeta$ for isotropic turbulence forced at wavenumbers $k_{\rm f}/k_1=5$ with $\Rm=10$ (squares)
and $k_{\rm f}/k_1=10$ with $\Rm=3.5$ (triangles).
Adapted from \cite{BRS08}.
}\label{ppk_kernel}\end{figure}

In \Fig{ppk_kernel} we show results for the Fourier transformed integral
kernels $\tilde\alpha(k)$ and $\tilde\eta_{\rm t}(k)$.
Both $\tilde\alpha$ and $\tilde\eta_{\rm t}$ decrease monotonously with
increasing $|k|$.
The two values of $\tilde\alpha$ for a given $k/k_{\rm f}$ but different
$k_{\rm f}/k_1$ and $\Rm$ are always very close together.
The functions $\tilde\alpha (k)$ and $\tilde\eta_{\rm t} (k)$
are well represented by Lorentzian fits of the form
\EQ
\tilde\alpha(k)={\alpha_0\over1+(k/k_{\rm f})^2} \, ,\quad
\tilde\eta_{\rm t}(k)={\eta_{\rm t0}\over1+(k/2k_{\rm f})^2} \, .
\label{KernelsTurb}
\EN
In \Fig{ppk_kernel} we show the kernels $\hat\alpha(\zeta)$ and
$\hat\eta_{\rm t} (\zeta)$ obtained numerically.
Observationally, similar results have been obtained by \cite{Abramenko}.

The results presented in \Fig{ppk_kernel}
show no noticeable dependencies on $\Rm$.
Although we have not performed any systematic survey in $\Rm$,
we interpret this as an extension of the above--mentioned results of Sur et al.\ (2008)
for $\alpha$ and $\eta_{\rm t}$ to the integral kernels $\hat\alpha$
and $\hat\eta_{\rm t}$.
Of course, this should also be checked with higher values of $\Rm$.
Particularly interesting would be a confirmation of different widths
for the profiles of $\hat\alpha$ and $\hat\eta_{\rm t}$.

The challenge in solar and stellar dynamo theory is nowadays not just the
understanding of the nature and origin of magnetic fields in observed stars
and in the Sun, but also the understanding of simulated dynamos.
Here we have a clear chance in achieving one-to-one agreement because
the magnetic Reynolds numbers are still manageable.
Only when such agreement has been achieved will we be able to address
in a meaningful way solar and stellar dynamos.

\acknowledgments

This research was supported in part by the European Research Council
under the AstroDyn Research Project 227952 and the Swedish Research
Council under the grants 621-2011-5076 and 2012-5797.
The computations have been carried out at the National Supercomputer
Centre in Ume{\aa} and at the Center for Parallel Computers at the
Royal Institute of Technology in Sweden.


\end{document}